\documentclass{article}

\usepackage{arxiv}

\usepackage[utf8]{inputenc} 
\usepackage[T1]{fontenc}    
\usepackage{url}            
\usepackage{booktabs}       
\usepackage{amsfonts}       
\usepackage{nicefrac}       
\usepackage{microtype}      
\usepackage[colorlinks = TRUE, linkcolor=blue,citecolor=blue,urlcolor=blue]{hyperref}
\usepackage{lipsum}         
\usepackage{doi}

\usepackage{amsthm,amsmath,amssymb,enumitem,multirow,colortbl,caption,subcaption,float,appendix}
\usepackage[authoryear]{natbib}
\usepackage{graphicx}

\usepackage{cleveref}       
\usepackage{comment}
\usepackage{algorithm, algpseudocode}

\newcommand{\bA} {\mathbf {A}}
\newcommand{\bB} {\mathbf {B}}

\newcommand{\be} {\mathbf {e}}
\newcommand{\bD} {\mathbf {D}}
\newcommand{\bG} {\mathbf {G}}
\newcommand{\bI} {\mathbf {I}}

\newcommand{\bP} {\mathbf {P}}

\newcommand{\br} {\mathbf {r}}
\newcommand{\bR} {\mathbf {R}}

\newcommand{\bU} {\mathbf {U}}
\newcommand{\bW} {\mathbf {W}}
\newcommand{\bV} {\mathbf {V}}

\newcommand{\bX} {\mathbf {X}}
\newcommand{\bx} {\mathbf {x}}
\newcommand{\bY} {\mathbf {Y}}
\newcommand{\by} {\mathbf {y}}
\newcommand{\bZ} {\mathbf {Z}}
\newcommand{\balpha} {\mbox{\boldmath $\alpha$}}
\newcommand{\bbeta} {\mbox{\boldmath $\beta$}}

\newcommand{\bmu} {\mbox{\boldmath $\mu$}}
\newcommand{\bgamma} {\mbox{\boldmath $\gamma$}}

\newcommand{\bone} {\mathbf {1}}
\newcommand{\bzero} {\mathbf {0}}
\def\T{{ \mathrm{\scriptscriptstyle T} }}

\usepackage{indentfirst}

\title{Regression Analysis of Elliptically Symmetric Directional Data}

\author{Zehao Yu \\
	Department of Statistics\\
	University of South Carolina\\
	Columbia, SC 29201 \\
	\texttt{zehaoy@email.sc.edu} \\
	\And
	Xianzheng Huang \\
	Department of Statistics\\
	University of South Carolina\\
	Columbia, SC 29201 \\
	\texttt{huang@stat.sc.edu}}

\date{}

\renewcommand{\shorttitle}{ESAG Regression}

\hypersetup{
	pdftitle={Regression Analysis of Elliptically Symmetric Directional Data},
	pdfsubject={statistics},
	pdfauthor={Zehao Yu and Xianzheng Huang},
	pdfkeywords={Angular Gaussian,  Hypersphere, Isotropy, Prediction region},
}

\numberwithin{equation}{section}
\theoremstyle{plain}

\begin{document}
\maketitle

\begin{abstract}
A comprehensive toolkit is developed for regression analysis of directional data based on a flexible class of angular Gaussian distributions. Informative testing procedures for isotropy and covariate effects on the directional response are proposed. Moreover, a prediction region that achieves the smallest volume in a class of ellipsoidal prediction regions of the same coverage probability is constructed. 
The efficacy of these inference procedures is demonstrated in simulation experiments. Finally, this new toolkit is used to analyze directional data originating from a hydrology study and a bioinformatics application. 
\end{abstract}

\keywords{Angular Gaussian \and Hypersphere \and Isotropy \and Prediction region}

\section{Introduction}
\label{s:intro}
Directional data naturally arise in many scientific disciplines, such as flight directions of migrating birds, the directions of wind and waves in the ocean, and geomagnetic field directions. These examples of directional data as the original form of observed data are typically of low dimensions. High dimensional directional data typically result from preprocessing high dimensional features collected in genetic study \citep{banerjee2005clustering}, computer vision \citep{ryali2013parcellation}, and text analysis \citep{ennajari2021combining}, among many other research fields. In these instances, the raw data vectors in some $d$-dimensional Euclidean space $\mathbb{R}^d$ are often normalized to lie on a hypersphere $\mathbb{S}^{d-1}=\{\by \in \mathbb{R}^d: \, \|\by\|=1\}$, where $\|\by\|$ denotes the Euclidean norm of $\by$.    

Regression analysis of directional data is relatively underdeveloped compared to regression analysis of response data in the linear (Euclidean) space. 
One of the most notable early developments of regression models for directional data is given by \citet{johnson1978some}, who formulated parametric models for the joint distribution of a circular response (i.e., $d=2$) and a linear covariate. Later,  \citet{presnell1998projected} introduced the spherically projected multivariate linear model based on the projected Gaussian distribution for the circular response with a mean direction depending on covariates linearly. Mimicking the least squares method in regression analysis for a linear response, \citet{lund1999least} proposed a least circular-distance method for regression analysis of a circular response. \citet{scealy2019scaled} proposed a transformation of the von Mises-Fisher distribution to study paleomagnetic data, following which they built regression models using the proposed directional distribution. \citet{paine2018elliptically} proposed the {\underline e}lliptically {\underline s}ymmetric {\underline a}ngular {\underline G}aussian distribution (ESAG), focusing on directional data on $\mathbb{S}^2$. In a follow-up work \citep{paine2020spherical}, the authors formulated regression models based on ESAG of low dimensions. 

The formulation of ESAG results from imposing constraints on the mean $\bmu$ and variance-covariance matrix $\bV$ of a multivariate Gaussian distribution $\mathcal{N}_d(\bmu, \bV)$ to resolve the identifiability issue. Such an identifiability issue emerges inevitably when normalizing a multivariate Gaussian vector to yield an angular Gaussian random variable, since two Gaussian vectors, $\bW$ and $c\bW$, are normalized to the same vector when $c>0$, yet they follow different Gaussian distributions whenever $c \ne 1$. For most angular Gaussian distributions, constraints are imposed on $\bV$ that often translate to stringent assumptions on the resultant directional distribution. The constraints on $\bmu$ and $\bV$ that lead to ESAG give rise to a probability density function (pdf) that does not involve a complicated normalization constant, and the resultant distribution remains flexible in that it is not limited to isotropic distributions. These are virtues of ESAG that make it stand out among many existing named directional distributions. Section~\ref{sec:models} provides a brief review of ESAG and its parameterization that facilitates regression analysis of directional data. We then address three inference problems that are theoretically and practically important in the context of directional distributions and regression analysis. More specifically, Section~\ref{sec:hypo} presents a novel diagnostic test to check isotropy. An isotropic distribution supported on a hypersphere is rotation-invariant, that is, rotating an isotropic random vector does not change its distribution. Isotropy is similar to symmetry for a distribution supported on a linear space, and thus testing isotropy is loosely parallel to testing the symmetry of a distribution. In Section~\ref{sec:covtest}, we propose methods for testing covariate-dependence of ESAG model parameters. Section~\ref{sec:simu} reports simulation studies for assessing the operating characteristics of the proposed testing procedures. 
 Section~\ref{sec:predict} provides prediction regions of the directional response.  We apply these new inference procedures to two real-life applications in Section~\ref{sec:realdata}. Section~\ref{sec:discuss} recapitulates the contributions of our study and points out some limitations of the proposed regression framework that motivate follow-up research.

\section{The ESAG regression model and likelihood-based inference}
\label{sec:models}
\subsection{The model and data}
A random variable $\bY$ supported on $\mathbb{S}^{d-1}$ follows an angular Gaussian distribution, $\text{AG}(\bmu, \, \bV)$, if $\bY=\bW/\|\bW\|$ with $\bW\sim \mathcal{N}_d(\bmu, \, \bV)$. The parameter $\bmu$ in $\text{AG}(\bmu, \bV)$ is the mean direction of $\bY$. To guarantee the identifiability of the distribution $\text{AG}(\bmu, \, \bV)$, assumptions on $(\bmu, \bV)$ are needed to avoid overparameterization. For example, \citet{presnell1998projected} assumed $\bV=\bI_d$ that leads to an isotropic directional distribution, where $\bI_d$ is the $d$-dimensional identity matrix. Less stringent assumptions are also considered, for example, in \citet{wang2013directional} where a sub-block of $\bV$ is assumed known. We adopt the ESAG distribution \citep{paine2018elliptically} resulting from imposing the following constraints that we refer to as ESAG constraints henceforth, 
$\bV\bmu = \bmu$ and $\mbox{det}(\bV) =1$, where $\mbox{det}(\bV)$ dentoes the determinant of $\bV$. These constraints leave more room for flexible modeling of $\bY$ than most previously considered constraints, at the price of creating a more complex constrained parameter space. We recently reparameterized ESAG by introducing constraint-free parameters $\bgamma\in \mathbb{R}^{(d-2)(d+1)/2}$ so that $\bV$ that satisfies ESAG constraints can be determined by $(\bmu, \bgamma)$ via an eigendecomposition \citep{yu2022elliptically}. Henceforth, we use $\bY \sim \text{ESAG}(\bmu, \, \bgamma)$ to refer to $\bY\sim \text{AG}(\bmu, \, \bV)$ with ESAG constraints imposed on $(\bmu, \bV)$.

The benefits of modeling ESAG via constraint-free parameters are at least twofold. First, maximum likelihood estimation of model parameters becomes more straightforward than directly estimating $(\bmu, \bV)$ subject to the nonlinear ESAG constraints. Second, a covariate-dependent ESAG can be easily formulated without introducing link functions to relate covariates to constrained model parameters as done in earlier regression models for directional responses \citep{lund1999least, scealy2011regression, scealy2017directional}. In this study, we consider an ESAG regression model specified by $\bY|\bX\sim \mbox{ESAG}(\bmu = \balpha_0+\bA_1 \bX, \, \bgamma = \bbeta_0+ \bB_1\bX)$, where $\bX = (X_{1},...,X_{q})^\top$ is the $q$-dimensional covariate vector, $\balpha_0$ is the intercept for modeling $\bmu$, $\bA_1 = [\balpha_1 \mid \ldots \mid \balpha_q]$ is the $d\times q$ matrix of regression coefficients representing covariates effects on $\bmu$, $\bbeta_0$ is the intercept parameter in $\bgamma$, and $\bB_1 = [\bbeta_1 \mid \ldots \mid \bbeta_q]$ is the $(d-2)(d+1)/2\times q$ matrix of covariates effects on $\bgamma$, in which $\balpha_k \in \mathbb{R}^d$ and $\bbeta_k \in \mathbb{R}^{(d-2)(d+1)/2}$, for $k=0,1,...,q$. 

Suppose the observed data include directional responses $\{\bY_1, \ldots, \bY_n\}$ from $n$ independent experimental units along with their covariates data 
$\{\bX_1, \ldots, \bX_n\}$. 
Similar to the treatment on covariates data in \citet{scealy2019scaled}, we standardize covariates data via $(X_{i,k} - X_{(1),k})/(X_{(n), k}-X_{(1), k})+1$, for $i=1, \ldots, n$, where $X_{(1), k}$ and $X_{(n), k}$ are the minimum and maximum order statistics corresponding to covariate $X_k$, for $k=1, \ldots, q$. The resultant standardized covariates data are more comparable in scale with the response of a unit Euclidean norm, which helps to stabilize the numerical implementation of maximum likelihood estimation without distorting the underlying association between the response and covariates. With a slight abuse of notation, we use $\{\bX_i\}_{i=1}^n$ to refer to the standardized covariates data. 

\subsection{Maximum likelihood estimation}
To parameterize $\bV$ in $\text{AG}(\bmu, \, \bV)$ to satisfy ESAG constraints, we introduced longitude and latitude angle parameters to specify eigenvectors of $\bV$ after $\bmu$ is specified. We showed that $\bgamma$ or a certain subvector of it being zero amounts to some latitude angles falling on the boundary of 0 or $\pi$ and some other latitude and longitude angles being non-identifiable \citep[see Appendix B in][for details]{yu2022elliptically}. This suggests violations of regularity conditions in the context of drawing likelihood-based inference for model parameters even though the parameter space of $\text{ESAG}(\bmu, \bgamma)$ is the entire real space $\mathbb{R}^{(d-1)(d+2)/2}$. The irregularity carries over to the ESAG regression model. As a result, maximum likelihood estimators (MLE) of some regression coefficients may converge in distribution to Gaussian at a slower rate than $\sqrt{n}$, or may not be asymptotically Gaussian, depending on where the true model parameters fall in the parameter space. Regardless, numerical implementation maximum likelihood estimation is straightforward under the current parameterization of ESAG, as demonstrated in our earlier work (and thus omitted here), and a simple resample-based bootstrap procedure can be used to quantify the uncertainty of the MLEs.  

When it comes to hypothesis testing, the conventional likelihood ratio test (LRT) is inadequate when regularity conditions are not satisfied because the asymptotic null distribution of a likelihood ratio (LR) statistic is no longer a $\chi^2$ \citep{chernoff1954distribution}. 
Most existing solutions to this complication with the LRT aim at estimating the exact distribution of LR or its limiting distribution under the null using some simulation-based methods, such as the method proposed by \citet{drton2009likelihood} and the approach developed in \citet{mitchell2019hypothesis}. Instead of using LR, we propose different test statistics that exploit unique properties of the ESAG distribution. These are elaborated in the next two sections, one focusing on tests for isotropy, and the other considering tests for covariate dependence of $\bmu$ and $\bgamma$. 

\section{Hypothesis testing for isotropy}
\label{sec:hypo}
If $\bY$ follows an isotropic distribution, then $\bR\bY$ and $\bY$ are identically distributed for any given $d \times d$ rotation matrix $\bR$. By the parameterization of $\bV$ via $\bgamma$, $\text{ESAG}(\bmu, \bgamma)$ is isotropic when $\bgamma=\bzero$, which gives $\bV = \bI_d$. Hence, testing isotropy is relevant to inferring correlations of the components in $\bW$, i.e., the pre-normalization version of $\bY$, and also relates to model selection between the more parsimonious isotropic ESAG and a generic ESAG distribution. In what follows, we propose a strategy for testing the null hypothesis $H^{(\bV)}_0: \bY \sim \text{ESAG}(\bmu, \bgamma=\bzero)$, where potential dependence of $\bmu$ on covariates $\bX$ is suppressed for notational simplicity. The proposed strategy is motivated by the properties of the MLE for the concentration parameter in the presence of model misspecification.

\subsection{Concentration estimation}
\label{sec:motivateRoC}
For $\text{ESAG}(\bmu, \bgamma)$, $\|\bmu\|$ quantifies the overall concentration of the distribution, with $\bV$ controlling the variation in different subspaces on the unit sphere. Visually, the shape of a data cloud from an isotropic $\text{ESAG}$ resembles a $(d-1)$-dimensional sphere, whereas the shape of a data cloud from an anisotropic ESAG is like a $(d-1)$-dimensional ellipsoid. Intuitively, when fitting an isotropic ESAG model to data from an anisotropic distribution, one essentially tries to find a ball that can most compactly contain an ellipsoid. To accomplish this, the radius of the ball tends to approach that of the longest axis of the ellipsoid, leading to a lower concentration of the fitted isotropic ESAG compared to the concentration of the true anisotropic distribution. In the context of model comparison, two ESAG distributions, $\text{ESAG}(\bmu_1, \bgamma_1=\bzero)$ and $\text{ESAG}(\bmu_2, \bgamma_2 \ne \bzero)$, are more alike when $\|\bmu_1\|<\|\bmu_2\|$ than when $\|\bmu_1\| \ge \|\bmu_2\|$. We demonstrate this phenomenon next by exploiting the properties of MLEs in the presence of model misspecification. 

Let $P$ denote a generic ESAG distribution with pdf $P(\bY; \bmu_a, \bgamma_a)$, which specifies the true data-generating mechanism. Let $Q$ denote an isotropic ESAG distribution with pdf $Q(\bY; \bmu)$. The Kullback–Leibler divergence of $Q$ from $P$ is defined as $D_{\text{KL}}(P\| Q; \bmu)=E_P[\log \{P(\bY; \bmu_a, \bgamma_a)/Q(\bY; \bmu)\}]$, where the subscript ``$P$'' signifies that the expectation is with respect to the distribution $P$. Under regularity conditions \citep{white1982maximum}, if one fits the model $Q$ to data from $P$, then the MLE for $\bmu$ converges in probability to $\bmu_0=\text{argmin}_{\bmu} D_{\text{KL}}(P\| Q; \bmu)=\text{argmax}_{\bmu}E_P\{\log 
 Q(\bY; \bmu)\}$. We show next that 
 $\|\bmu_0\|\le \|\bmu_a\|$, or, equivalently, in the presence of model misspecification (i.e., $P\ne Q$), $E_P\{\log 
 Q(\bY; \bmu)\}$ is maximized when the ratio of concentrations (RoC) $\|\bmu_a\|/\|\bmu_0\|$ exceeds one. To highlight the concentration, we view $\bmu_a=c_a \bR_a \bmu^*$ and $\bmu_0=c_0 \bR_0 \bmu^*$ for some rotation matrices, $\bR_a$ and $\bR_0$, and some positive constants, $c_a$ and $c_0$, where $\bmu^*$ is a unit vector. In other words, $\bmu_a$ and $\bmu_0$ may differ in concentration, quantified by $c_a$ and $c_0$ respectively, or differ in orientation when $\bR_a \ne \bR_0$. Using this factorization of the mean direction parameter, we have $\|\bmu_a\|/\|\bmu_0\|=c_a/c_0$ since $\|\bR_a \bmu^*\|/\|\bR_0 \bmu^*\|=1$. Now we re-express the density $P(\cdot; \bmu_a, \bgamma_a)$ as $P(\cdot; c_a, \bR_a, \bgamma_a)$, and similarly write the density $Q(\cdot; \bmu)$ as $Q(\cdot; c, \bR)$, where the dependence of these distributions on $\bmu^*$ is suppressed because the value of $\bmu^*$ remains the same for all ESAG distributions under this formulation of the mean direction parameter. Without loss of generality, let $\bmu^*=\bmu_a/\|\bmu_a\|$. With this choice of $\bmu^*$, we have $c_a=\|\bmu_a\|$ and $\bR_a=\bI_d$. Fitting $Q$ to data from $P$ now amounts to, in limit as $n\to \infty$, maximizing $E_P\{\log 
 Q(\bY; c, \bR)\}$ with respect to $(c, \bR)$, which cannot be done analytically but can be simulated using large samples.  

To simulate this maximization problem, we generate a random sample of size $n=10^4$ from $P(\cdot; c_a, \bR_a, \bgamma_a)$ for a $\bmu_a$ we specify, and $\bgamma_a$ taking one of the following three values, $\bgamma^{(1)}=\bzero$, non-zero $\bgamma^{(2)}$ and $\bgamma^{(3)}$, with the first value creating a scenario where $P=Q$, and the latter two creating increasing degree of anisotropy in $P$. We then use the log-likelihood function $\ell(c, \bR)=n^{-1}\sum_{i=1}^n \log Q(\bY_i; c, \bR)$ as an empirical version of $E_P\{\log 
 Q(\bY; c, \bR)\}$ to demonstrate that $c_a/c^*>1$ when $\bgamma_a \ne \bzero$, where $c^*=\text{argmax}_{c>0} \ell (c, \bR^*)$ for some arbitrary rotation matrix  $\bR^*$. For concreteness, we consider three values for $\bR^*$ given by $\bR^{(1)}=\bR_a$, $\bR^{(2)}\ne \bR_a$, and $\bR^{(3)}$ that deviates from  $\bR_a$ further than $\bR^{(2)}$ does.  

\begin{figure}[htp]
\centering
\includegraphics[width=5in]{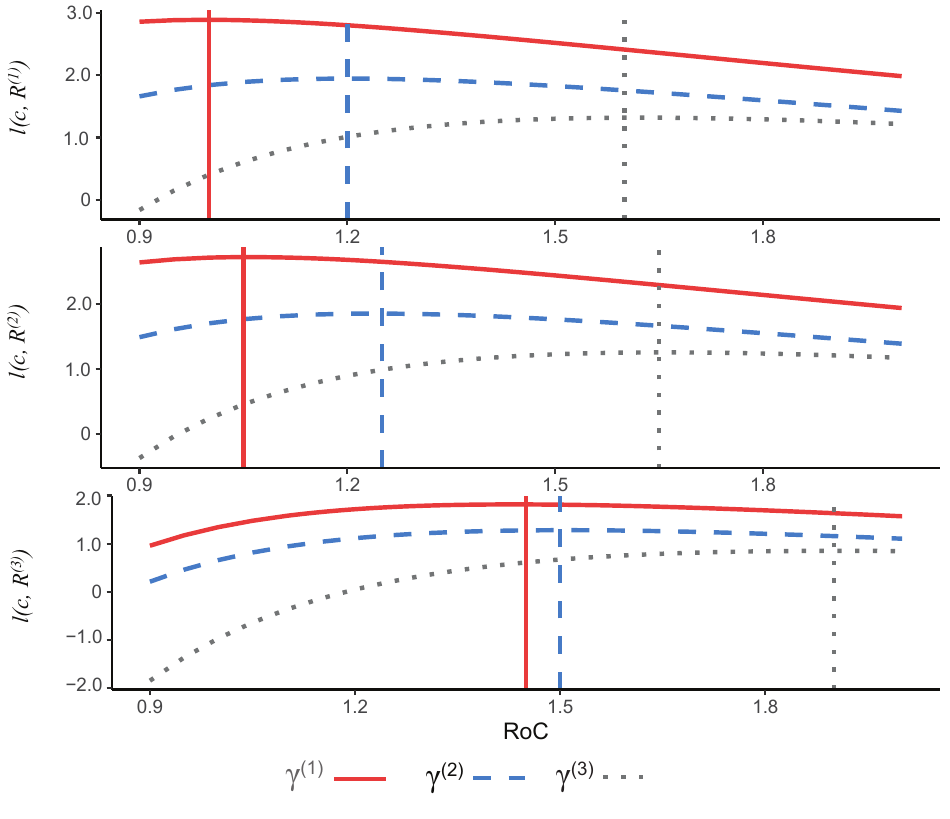}
\caption{The empirical version of  $E_P\{\log Q(\bY; c,\bR)\}$, $\ell(c, \bR)$, based on a random sample of size $n=10^4$ from an isotropic ESAG (red lines), a mildly anisotropic ESAG with $\bgamma=\bgamma^{(2)}\ne 
 \bzero$ (blue dashed lines), and an anisotropic ESAG with $\bgamma=\bgamma^{(3)}$ deviating from $ 
 \bzero$ even further (gray dotted lines), versus RoC when $\bR$ is set at $\bR^{(1)}=\bI_d$ (top panel), $\bR^{(2)}\ne \bI_d$ (middle panel), and $\bR^{(3)}$ that deviates from $\bI_d$ even more (bottom). Vertical lines mark the value of RoC where the corresponding function $\ell(c, \bR)$ is maximized.}
\label{fig:ElogQ}
\end{figure}

The top panel of Figure~\ref{fig:ElogQ} depicts $\ell(c, \bR^{(1)})$ as a function of $\text{RoC}=c_a/c$ when the data-generating mechanism $P$ has $\bgamma_a$ set at $\bgamma^{(1)}=\bzero$ (isotropy), $\bgamma^{(2)}\ne \bzero$ (mild anisotropy), and $\bgamma^{(3)}$ that deviates from zero even further (severe anisotropy), respectively. With $\bR^{(1)}=\bR_a$, the mean directions of $P$ and $Q$ have the same orientation. When $P$ is isotropic, $E_P\{\log Q(\bY; c, \bR^{(1)})\}$ is expected to be maximized at $c^*=c_a$,  resulting in $D_{\text{KL}}(P\|Q; \bmu_0)=0$. This is indeed (empirically) justified by the curve of $\ell(c; \bR^{(1)})$ that reaches its peak at around $\text{RoC}=c_a/c^*=1$. Once $P$ exhibits anisotropy by having $\bgamma_a$ deviating from $\bzero$, one witnesses a drop in the likelihood $\ell(c, \bR^{(1)})$, which is maximized at some RoC that exceeds 1, indicating that $c^*<c_a$. The inflation in RoC, i.e., the attenuation in $c^*$, becomes more substantial as $\bgamma_a$ deviates from $\bzero$ further. This implies that misspecifying $\bgamma$ in the ESAG distribution by assuming isotropy can be manifested in a larger-than-1 RoC. The bottom two panels in Figure~\ref{fig:ElogQ} show  $\ell(c, \bR^{(2)})$ and $\ell(c, \bR^{(3)})$ versus RoC, where all the previously observed phenomena for $\ell(c, \bR^{(1)})$ remain except for that, even with $\bgamma_a$ set at $\bzero$, $\ell(c, \bR^*)$ is also maximized when RoC is larger than 1.

Comparing the three panels in Figure~\ref{fig:ElogQ} reveals a clear trend of RoC increasing as model misspecification becomes more severe by having $\bgamma_a$ further away from zero or having the orientation of $\bmu_a$ mismatch more with the orientation of $\bmu_0$. The latter observation suggests that RoC can be used to test assumptions regarding $\bmu$ as well, which is a point we come back to in a later section on testing assumptions on the mean direction parameter.

\subsection{Testing isotropy based on concentration estimation}
\label{sec:testRoC}
Inspired by the above findings regarding concentration estimation, we propose the statistic for testing isotropy defined by 
\begin{equation}
 \text{RoC} = \frac{1}{n}\sum_{i=1}^n\frac{\|\hat\bmu_{ai}\|}{\|\hat\bmu_{0i}\|},\label{eq:testiso}
\end{equation}
where $\hat\bmu_{0i}$ is the restricted MLE of the mean direction of $\bY_i$ given $\bX_i$ under $H_0^{(\bV)}$, and $\hat\bmu_{ai}$ is the unrestricted MLE under the alternative hypothesis that allows anisotropy. If the true data-generating mechanism is consistent with $H_0^{(\bV)}$, then RoC is expected to be close to one; otherwise, RoC tends to be larger than one. 

 Algorithm~\ref{alg:RoC} below gives the parametric bootstrap procedure to estimate the $p$-value associated with RoC to assess its statistical significance. The goal of the bootstrap procedure is to estimate the null distribution of RoC by simulating realizations of RoC under the null. To this end, we repeatedly compute RoC based on data generated from an isotropic ESAG distribution $\hat Q(\cdot; \hat \bmu_{0i})$ for the $i$-th experimental unit, for $i=1, \ldots, n$. This distribution is an estimate of $Q$ that is closest to the unknown true model $P$ for each experimental unit.  An estimated $p$-value can then be obtained by comparing the RoC computed based on the raw data with the simulated RoC's. As seen here and in the testing procedures proposed later for other purposes, the distribution of a test statistic under any hypothesized ESAG model can be easily approximated via parametric bootstrap because it is straightforward to simulate data from any ESAG distribution, which is yet another virtue of the ESAG distribution family and the constrain-free parameterization. 
\begin{algorithm}[h]
\footnotesize
  \caption{Hypothesis testing for isotropy based on RoC defined in \eqref{eq:testiso}\label{alg:RoC}}
  \begin{algorithmic}[1]
    \Procedure{Compute RoC based on the observed data}{}
        \State Input data $\lbrace(\bY_i,\bX_i)\rbrace_{i=1}^n$, find the restricted MLE $\hat{\bmu}_{0i}$ under $H_0^{(\bV)}$ and the unrestricted MLE $\hat{\bmu}_{ai}$, $\hat{\bgamma}_{ai}$, for $i=1, \ldots, n$.
        \State Compute the test statistic $\mbox{RoC} = (1/n)\sum_{i=1}^n \|\hat{\bmu}_{ai}||/ \|\hat{\bmu}_{0i}||$. 
    \EndProcedure

    \Procedure{Bootstrap procedure to estimate the null distribution of RoC}{}
        \State Set $B$ = number of bootstraps
        \State Initiate $s = 0$
        \For{$b$ in $1,...,B$}
        \State Generate the $b$-th bootstrap sample $\{\bY_i^{(b)}\}_{i=1}^n$, where $\bY_i^{(b)}|\bX_i\sim\mbox{ESAG}(\hat{\bmu}_{0i}, \, \bgamma_i=\bzero )$, for $i = 1, ..., n$.
        \State Repeat Steps 2--3 using data $\lbrace(\bY_i^{(b)}, \bX_i)\rbrace_{i=1}^n$. Denote the resultant value of $\mbox{RoC}$ as $\mbox{RoC}^{(b)}$. 
        \State \textbf{if} $ \mbox{RoC}^{(b)}>\mbox{RoC}$ \textbf{then}  $s = s + 1$ 
      \EndFor
      \State  Output $s/B$ as an estimated $p$-value associated with RoC from Step 3.    \EndProcedure
  \end{algorithmic}
\end{algorithm}

\section{Tests for covariates effects}
\label{sec:covtest}
\subsection{Testing covariates dependence of $\bmu$}
\label{sec:mucov}
For a directional response, a practically interesting question is whether or not its mean direction depends on covariates. For concreteness, let us consider testing the null $H_0^{(\bmu)}:\, \bY|\bX\sim \text{ESAG}(\bmu=\balpha_0, \, \bgamma=\bbeta_0+\bB_1 \bX)$ versus the alternative $H_1: \, \bY|\bX\sim \text{ESAG}(\bmu=\balpha_0+\bA_1  \bX, \, \bgamma=\bbeta_0+\bB_1 \bX)$. If the alternative is true with $\bA_1\ne \bzero$, the fitted $\bmu$ under the null, denoted by $\hat \bmu_0$, is expected to differ from the fitted value that allows covariates dependence of $\bmu$, denoted by $\hat\bmu_a$. The difference can lie in their norms, i.e., concentrations, or in their directions. This motivates the following test statistic that captures both sources of discrepancies, 
\begin{align}
D =  \frac{1}{n}\sum_{i=1}^n \left( 2 - \frac{\hat\bmu_{0i}^\top \hat\bmu_{ai}}{\|\hat \bmu_{0i} \| \| \hat \bmu_{ai} \|}\right) \frac{\|\hat \bmu_{ai} \|}{\| \hat\bmu_{0i} \|}, \label{eq:testcovD}
\end{align}
where, for the $i$-th data point, $\hat \bmu_{0i}$ is the restricted MLE obtained under the null that assumes covariate-independent $\bmu$, and $\hat \bmu_{ai}$ is the unrestricted MLE obtained under the alternative. In \eqref{eq:testcovD},  $\hat \bmu_{0i}^\top \hat \bmu_{ai}/(\|\hat\bmu_{0i} \| \| \hat\bmu_{ai} \|)$ is known as the cosine similarity between two vectors, $\hat \bmu_{0i}$ and $\hat \bmu_{ai}$, which is equal to 1 if they have the same direction, and is equal to $-1$ if the directions are opposite. Hence the first factor in the summand in \eqref{eq:testcovD} quantifies the dissimilarity in direction between $\hat \bmu_{0i}$ and $\hat \bmu_{ai}$.  The second factor of the summand in \eqref{eq:testcovD} contrasts the concentrations of the two estimates for $\bmu$ as in RoC.  By construction, under the null $H_0^{(\bmu)}$, $D$ is expected to be close to 1; and a realization of $D$ that is much larger than 1 can imply the observed data coming from a model that violates of the null. 

 Algorithm~\ref{alg:D} provides detailed steps for implementing the test based on the newly proposed test statistic, where we again use a parametric bootstrap procedure to estimate the $p$-value associated with $D$. 
\begin{algorithm}[h!]
\footnotesize
  \caption{Hypothesis testing regarding $\bmu$ based on $D$ defined in \eqref{eq:testcovD}\label{alg:D}}
  \begin{algorithmic}[1]
    \Procedure{Compute $D$ based on the observed data}{}
        \State Input data $\lbrace(\bY_i,\bX_i)\rbrace_{i=1}^n$, find the restricted MLEs $\hat \bmu_{0i}$ and $\hat \bgamma_{0i}$, and the unrestricted MLEs $\hat \bmu_{1i}$ and $\hat \bgamma_{1i}$, for $i=1, \ldots, n$.
        \State Compute $D = (1/n)\sum_{i = 1}^n [\{ 2 - (\hat \bmu_{0i}^\top  \hat\bmu_{ia})/(\|\hat\bmu_{0i} \| \|\hat \bmu_{ai} \|)\} (\|\hat\bmu_{ai} \|/\| \hat\bmu_{0i} \|) ]$.
    \EndProcedure
    \Procedure{Bootstrap procedure to estimate the null distribution of $D$}{}
        \State Set $B$ = number of bootstraps
        \State Initiate $s = 0$
        \For{$b$ in $1,...,B$}
        \State Generate the $b$-th bootstrap sample $\{\bY_i^{(b)}\}_{i=1}^n$, where $\bY_i^{(b)}|\bX_i\sim\mbox{ESAG}(\hat \bmu_{0i}, \, \hat\bgamma_{0i} )$, for $i = 1, ..., n$.
        \State Repeat Steps 2--3 using data $\lbrace(\bY_i^{(b)}, \bX_i)\rbrace_{i=1}^n$. Denote the resultant test statistic as $D^{(b)}$. 
        \State \textbf{if} $D^{(b)}>D$ \textbf{then}  $s = s + 1$ 
      \EndFor
      \State  Output $s/B$ as an estimated $p$-value associated with $D$ from Step 3. 
    \EndProcedure
  \end{algorithmic}
\end{algorithm}

As indicated in Section~\ref{sec:motivateRoC}, RoC can be used to test hypotheses about $\bmu$, such as testing covariate dependence of it by adopting Algorithm~\ref{alg:RoC} with the restricted MLEs for $\bmu$ and $\bgamma$ obtained under the current null $H_0^{(\bmu)}$. Moreover, because $D$ incorporates information regarding direction comparison between two fitted values of $\bmu$ besides information relating to concentration comparison that RoC focuses on, one can combine the two test statistics to gain more insight into the underlying data-generating mechanism. If $D$ is significantly larger than RoC when testing covariate dependence of $\bmu$, one may interpret it as data evidence for the direction of $\bmu$ depending on some covariate. Having $D$ close to RoC can imply that the direction of $\bmu$ may not be dependent on covariates, although its norm may depend on covariates. This exemplifies the versatility and additional insight our proposed test statistics can offer when compared with LR.

\subsection{Testing covariates dependence of $\bgamma$ and beyond}
\label{sec:testgammacov}
Unique to our parameterization of $\text{ESAG}(\bmu, \bgamma)$, parameters in $\bgamma$ control variation of the distribution in different subspaces on the hypersphere besides (an)isotropy. It is thus of interest to test if such distributional features depend on covariates. For instance, one may consider testing the null $H^{(\bgamma)}_0: \, \bY|\bX\sim \text{ESAG}(\bmu=\balpha_0+\bA_1 \bX, \, \bgamma=\bbeta_0)$ versus the alternative $H_1: \, \bY|\bX\sim \text{ESAG}(\bmu=\balpha_0+\bA_1 \bX, \, \bgamma=\bbeta_0+\bB_1 \bX)$. Because $\bgamma$ as a whole relates to (an)isotropy of the distribution, RoC that is initially proposed for testing isotropy has its natural appeal for testing hypotheses about $\bgamma$. When a violation of $H_0^{(\bgamma)}$ adversely affects inferences for $\bmu$, the test statistic $D$ designed for testing assumptions on $\bmu$ also has the potential to detect covariates dependence of $\bgamma$. With the restricted MLEs $\hat \bmu_{0i}$ and $\hat \bgamma_{0i}$ now reflecting $H^{(\bgamma)}_0$ used in Algorithm~\ref{alg:RoC} or Algorithm~\ref{alg:D}, one can carry out the
test based on RoC or $D$ to test $H_0^{(\bgamma)}$. 

Fixing $H_1$ at the above saturated ESAG model, to test other null hypotheses, say,  $\balpha_k =\bzero$ for a given  $k\in \{1, \ldots, q\}$, Roc and $D$ can be used with the restricted MLEs in Algorithms~\ref{alg:RoC} and \ref{alg:D} revised accordingly to reflect the specific null hypothesis under consideration. Even if one adopts an angular Gaussian distribution that is not ESAG, as long as the mean vector $\bmu$ has the same interpretations as that in ESAG($\bmu, \bgamma)$, RoC and $D$ remain meaningful statistics for testing assumptions on $\bmu$ or other model assumptions that inferences for $\bmu$ are sensitive to. One simply needs to revise the bootstrap procedures to adapt to the assumed angular Gaussian distribution.

Lastly, RoC and $D$ depend on both the restricted and unrestricted MLEs of model parameters, which in turn add to the computational burden in   Algorithms~\ref{alg:RoC} and \ref{alg:D} where these MLEs are obtained based on each bootstrap sample. We thus propose yet another testing strategy that only requires computing the restricted MLEs that is based on a second moment estimation, with the test statistic given by 
\begin{align}
    M & = \left\|\frac{1}{n}\sum_{i=1}^n \left \{   \bY_i^2 - \widehat{\textbf{E}_0\left(\bY_i^2\right)}  \right\} \right\|, \label{eq:testIsoM}
\end{align}
where $\bY^2_i$ is the element-wise quantity square of $\bY_i$, and $\widehat{\textbf{E}_0\left(\bY_i^2\right)}$ is an empirical mean of $\bY^2$ given $\bX=\bX_i$ computed using a random sample simulated from an estimated null model $\hat Q(\cdot; \hat\bmu_{0i})$. Unlike RoC and $D$, the construction of $M$ is not motivated by (and thus does not target at testing) a particular aspect of the model specification; instead, $M$ can serve as an overall goodness-of-fit test statistic. By construction, in the absence of model misspecification, $M$ is expected to be close to zero, and a larger $M$ serves as data evidence of a poor fit of a null model for the observed data. As an example, Algorithm~\ref{alg:Mome} below gives the algorithm for using $M$ to test the null model that assumes an isotropic ESAG, with an estimated $p$-value obtained via parametric bootstrap as an output. 
\begin{algorithm}[h!]
\footnotesize
  \caption{Hypothesis testing for isotropy based on $M$ defined in \eqref{eq:testIsoM}\label{alg:Mome}}
  \begin{algorithmic}[1]
    \Procedure{Compute $M$ based on the observed data}{}
        \State Input data $\lbrace(\bY_i,\bX_i)\rbrace_{i=1}^n$, find the restricted MLE $\hat{\bmu}_{0i}$, for $i=1, \ldots, n$.
        \State For $i=1, \ldots, n$, generate $\{\tilde \bY_{i,m}\}_{m=1}^{10^4}$ from $\text{ESAG}(\hat \bmu_{0i}, \bgamma_i=\bzero)$, compute   $\widehat{\textbf{E}_0(\bY_i^2)}=10^{-4}\sum_{m=1}^{10^4} \tilde\bY_{i,m}^2$.
        \State Compute $M = \| (1/n)\sum_{i=1}^n \{ \bY_i^2 - \widehat{\textbf{E}_0(\bY_i^2)}  \} \|$.
    \EndProcedure
    \Procedure{Bootstrap procedure to estimate the null distribution of $M$}{}
        \State Set $B$ = number of bootstraps
        \State Initiate $s = 0$
        \For{$b$ in $1,...,B$}
        \State Generate the $b$-th bootstrap sample $\{\bY_i^{(b)}\}_{i=1}^n$, where $\bY_i^{(b)}|\bX_i\sim\mbox{ESAG}(\bmu_i =\hat{\bmu}_{0i} , \, \bgamma_i = \bzero)$, for $i = 1, ..., n$.
        \State Repeat Steps 2--4 using data $\lbrace(\bY_i^{(b)}, \bX_i)\rbrace_{i=1}^n$. Denote the resultant test statistic as $M^{(b)}$. 
        \State \textbf{if} $ M^{(b)}>M$ \textbf{then}  $s = s + 1$ 
      \EndFor
      \State Output $s/B$ as an estimated $p$-value associated with $M$ from Step 4. \EndProcedure
  \end{algorithmic}
\end{algorithm}

\section{Simulation study}
\label{sec:simu}
\subsection{Design of simulation experiments}
We are now in the position to study empirically operating characteristics of the proposed testing procedures for testing 
 $H_0^{(\bV)}$, $H_0^{(\bmu)}$, and $H_0^{(\bgamma)}$ versus the alternative $H_1: \, \bY|\bX\sim \text{ESAG}(\bmu=\balpha_0+\bA_1 \bX, \, \bgamma=\bbeta_0+\bB_1 \bX)$. To this end, we design several data-generating mechanisms (DGM) for each null hypothesis. A random sample of size $n\in \{200, 400, 800\}$ is generated according to each DGM, based on which the proposed test statistics and their estimated $p$-values are computed following Algorithms~\ref{alg:RoC}--\ref{alg:Mome} with $B=300$.  As a benchmark testing procedure to compare with ours, we also test each null using LR, with the corresponding $p$-value estimated via parametric bootstrap, as opposed to assuming a $\chi^2$ null distribution for LR as in \citet{paine2020spherical}.  
 This experiment is repeated 200 times at each simulation setting specified by the null hypothesis, DGM, and the level of $n$. Common in all settings, we consider one covariate, with $n$ realizations $\{X'_i\}_{i=1}^n$ generated from $N(0, 1)$, followed by standardization via $X_i=(X'_i-X'_{(1)})/(X'_{(n)}-X'_{(1)})+1$, for $i=1, \ldots, n$. Given the covariate data $\{X_i\}_{i=1}^n$, response data $\{\bY_i\}_{i=1}^n$ are generated according to a DGM specified in Table~\ref{t:DGMs}.
\begin{table}[h!]
\centering
\caption{\label{t:DGMs}Data-generating mechanisms (DGM)   designed for testing each considered null hypothesis regarding $\text{ESAG}(\bmu, \bgamma)$, along with values of model parameters in these DGMs}
\begin{tabular}{ll}
\hline
Null hypothesis   & ESAG data-generating mechanism  \\ 
\hline \\
\multirow{3}{*}{$H_0^{(\bV)}:\, \bmu=\balpha_0+\balpha_1 X, \, \bgamma=\bzero$}  & $\text{DGM}^{(\bV)}_0: \, \bmu = \balpha_0^* +\balpha_1^* X_, \, \bgamma = \bzero$ \\
  & $\text{DGM}^{(\bV)}_1: \, \bmu = \balpha_0^* +\balpha_1^* X, \, \bgamma = \bbeta_{0r}^*$ \\ 
  & $\text{DGM}^{(\bV)}_2: \, \bmu = \balpha_0^* +\balpha_1^* X, \, \bgamma = \bbeta_{0r}^* + \bbeta_{1r}^*X$ \\
  \cline{2-2} \\
\multirow{4}{*}{$H_0^{(\bmu)}:\, \bmu=\balpha_0, \, \bgamma=\bbeta_0+\bbeta_1 X$} & $\text{DGM}^{(\bmu)}_0: \,  \bmu=\balpha^*_0, \, \bgamma=\bzero$ \\
& $\text{DGM}^{(\bmu)}_1: \,  \bmu=\balpha^*_0+\balpha^*_{1r}X, \, \bgamma=\bzero$ \\
& $\text{DGM}^{(\bmu)}_2: \,  \bmu=\balpha^*_0+\balpha^*_{1r}X, \, \bgamma=\bbeta_0^*$ \\
& $\text{DGM}^{(\bmu)}_3: \,  \bmu=\balpha^*_0+\balpha^*_{1r}X, \, \bgamma=\bbeta_0^*+\bbeta_1^* X$ \\
\cline{2-2} \\
\multirow{3}{*}{$H_0^{(\bgamma)}:\, \bmu=\balpha_0+\balpha_1 X, \, \bgamma=\bbeta_0$} & $\text{DGM}^{(\bgamma)}_0: \,  \bmu=\balpha^*_0+\balpha_1^* X, \, \bgamma=\bzero$ \\
& $\text{DGM}^{(\bgamma)}_1: \,  \bmu=\balpha^*_0+\balpha^*_1X, \, \bgamma=\bbeta_{1r}^* X$ \\
& $\text{DGM}^{(\bgamma)}_2: \,  \bmu=\balpha^*_0+\balpha^*_1X, \, \bgamma=\bbeta_0^*+\bbeta_{1r}^* X$ \\
\hline \\
\multirow{3}{*}{Values of model parameters} & $\balpha_0^*=(2,-5,3,5)^\top$,  \\
& $\balpha_1^*=(2,1,2,1)^\top,\,\balpha_{1r}^*= \displaystyle{\frac{r}{2}\bone_4}$, \\
& $\bbeta_0^*=(3,5,-3,-4,2)^\top, \, \bbeta_{0r}^*=\displaystyle{\frac{r}{\sqrt{5}}\bone_5}$ \\
& $\bbeta_1^*=(4,2,5,-2,3)^\top,\, \bbeta_{1r}^*=\displaystyle{\frac{r}{\sqrt{5}}\bone_5}$ \\
  \hline
\end{tabular}
\end{table}

As one can see in Table~\ref{t:DGMs}, for each considered null hypothesis, we include a DGM matching the null. This allows for inspecting the size of a test. For each considered null, we also design several DGMs with increasing model complexity compared to the null. The values of some regression coefficients depend on a quantity $r$ that we vary in the simulation to control the severity of model misspecification under a null, with a larger $r$ leading to a more pronounced deviation of the DGM from a null. This allows for monitoring the power of a test as the true model deviates from the null model further. 

The metric we record in the simulation study is the relative frequency of a considered test rejecting the current null across 200 Monte Carlo replicates at a pre-specified significance level. In what follows, we present these rejection rates associated with different tests for testing each of the three null hypotheses tabulated in Table~\ref{t:DGMs}.  

\subsection{Simulation results}
\begin{figure}[h!]
\begin{subfigure}{0.80\linewidth}
		\centering
    \includegraphics[width=6in,height=2in]{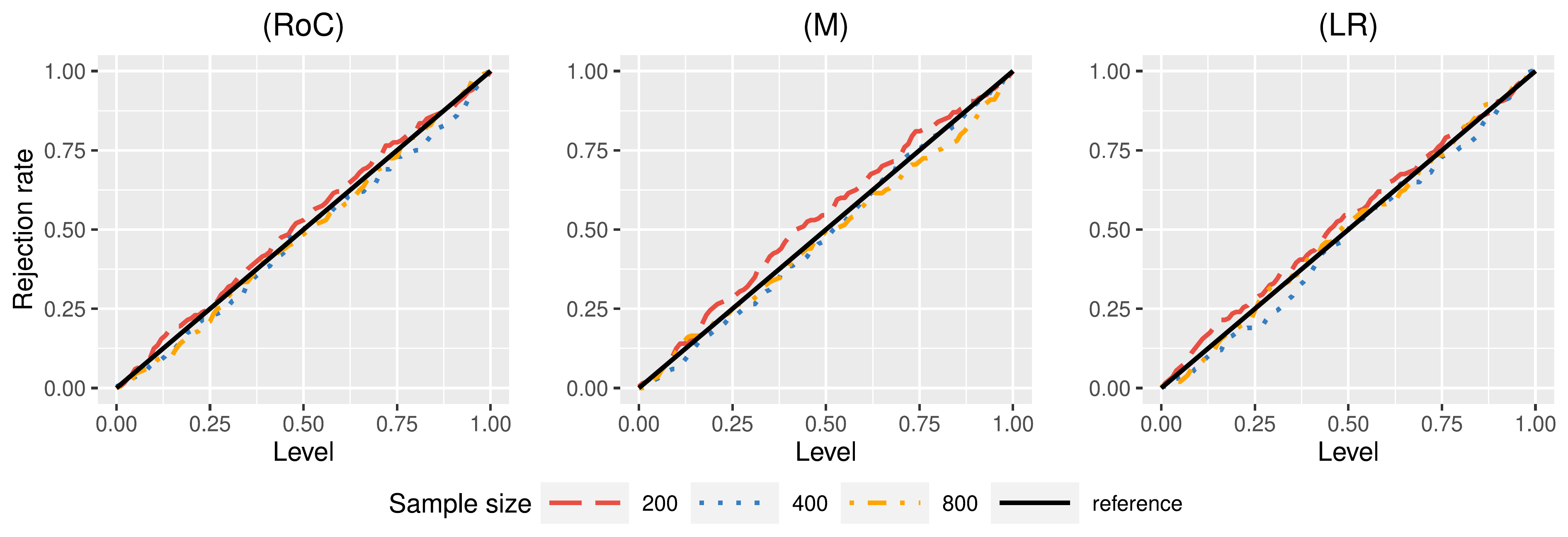}
\end{subfigure}
\begin{subfigure}{0.80\linewidth}
    \centering
    \includegraphics[width=6in,height=2in]{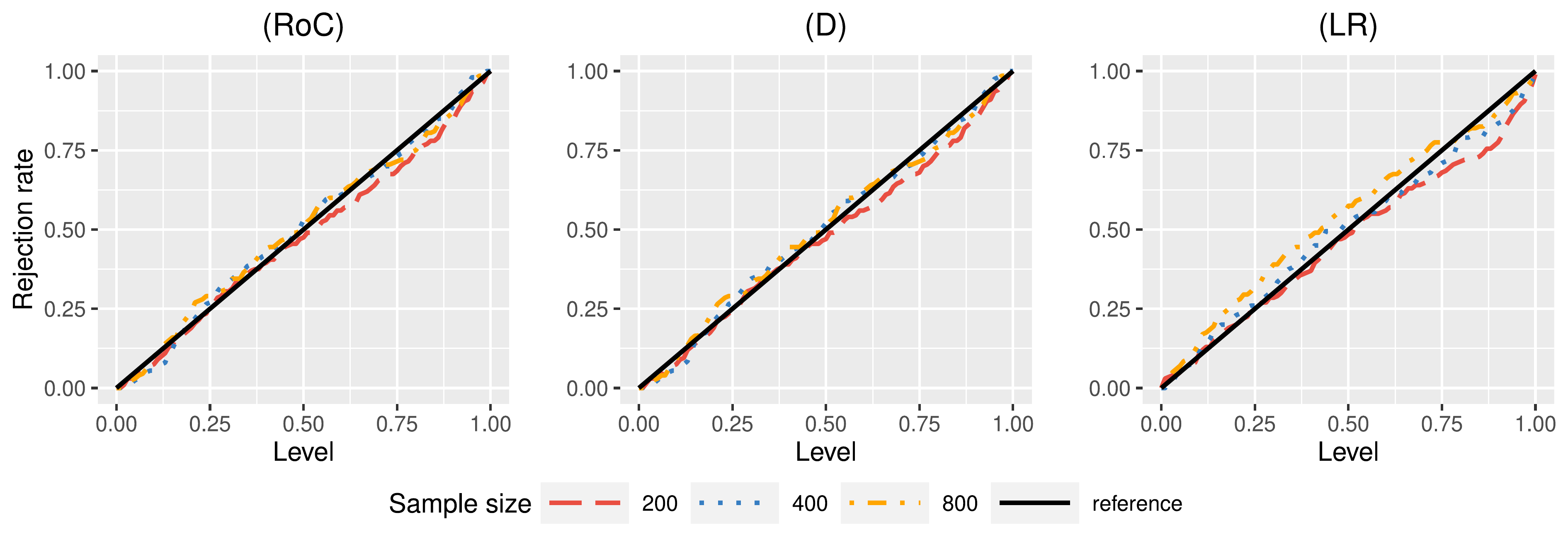}
\end{subfigure}
\begin{subfigure}{0.80\linewidth}
    \centering
    \includegraphics[width=6in,height=2in]{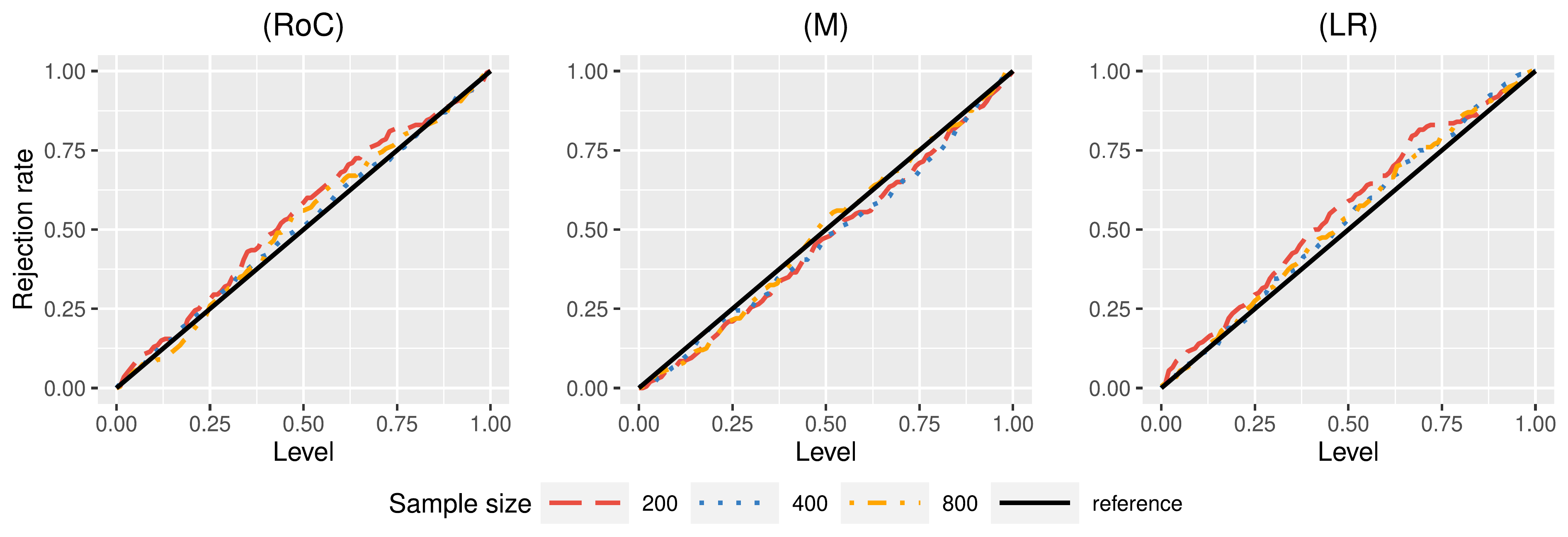}
\end{subfigure}
\caption{\label{fig:testsize}Rejection rates of tests for testing isotropy, $H_0^{(\bV)}$ (top row), covariate dependence of $\bmu$, $H_0^{(\bmu)}$ (middle row), and covariate dependence of $\bgamma$, $H_0^{(\bgamma)}$ (bottom row) when $n=200$ (red dashed lines), 400 (blue dotted lines), and 800 (orange dash-dotted lines). The black solid lines are the $45^\circ$ reference line.}
\end{figure}

Figure \ref{fig:testsize} provides the rejection rates of RoC, $D$ or $M$, and LR versus the nominal significance level based on data generated from an ESAG regression model consistent with a null hypothesis in Table~\ref{t:DGMs}. Focusing on the lower range of the nominal level such as 0.01 and 0.05, we conclude well-controlled sizes of all proposed tests, whereas the size of LRT may be subject to slight inflation, especially when testing covariate dependence of model parameters. This can be where the size of LRT fails to approach the nominal level asymptotically even when its $p$-value is estimated by the conventional parametric bootstrap, which is a phenomenon described in \cite{drton2011quantifying}. One shall thus interpret the empirical power of LRT with caution. For this reason, we omit to report the empirical power of LRT for testing covariate dependence. 

Table~\ref{table:dependXpower} presents the empirical power of various tests for testing each of the three null hypotheses at a significance level of 0.05 based on data from different true ESAG models. When using RoC, $M$, and LR to test isotropy, the three tests are comparable in their power to detect anisotropy, with the power increasing steadily as $n$ grows bigger or as the true value of $\bgamma$ deviates from zero further (by having a larger $r$). Having a covariate-dependent $\bgamma$ in the true regression model also enhances the power of these tests, although $M$ appears to be somewhat less powerful than RoC in this scenario. 

According to Table~\ref{table:dependXpower}, the tests based on RoC and $D$ enjoy higher power to detect covariate dependence of $\bmu$ when the true model also has a covariate-dependent $\bgamma$ (as in $\text{DGM}^{(\bmu)}_3$) than when it has an intercept-only model for $\bgamma$ (as in $\text{DGM}^{(\bmu)}_2$). Noting that obtaining the unrestricted MLE for $\bgamma$ using data from $\text{DGM}^{(\bmu)}_2$ creates an irregular maximum likelihood estimation, but the same estimation using data from $\text{DGM}^{(\bmu)}_3$ is a regular case, we believe that having irregular MLEs for model parameters can compromise the power of RoC and $D$.      
When testing $H_0^{(\bgamma)}$, the power of the proposed tests does not increase as quickly as when testing $H_0^{(\bmu)}$ when $n$ increases or when the covariate dependence becomes stronger. We conjecture that, once we allow $\bmu$ to depend on covariates, inferences for the concentration are less sensitive to the assumption of covariate-independent for $\bgamma$, and thus RoC and $D$ may lack high power to detect the dependence of $\bgamma$ on covariates unless when the dependence is very strong. 

The moment-based test using $M$ is much less powerful than the RoC test and the test based on $D$ for testing covariate dependence of model parameters. By solely focusing on the fit for the mean of $\bY^2$, the power $M$ to detect model misspecification heavily hinges on the impact of the misspecification on second-moment estimation. The observed phenomenon suggests some level of robustness of the second-moment estimation to covariate dependence of ESAG model parameters. In additional simulation study not reported here where we generate covariate data from different distributions, we observe that likelihood-based estimation of $E(\bY^2)$ is more sensitive to violation of $H_0^{(\bmu)}$ or $H_0^{(\bgamma)}$ when the covariate distribution is skewed, and, consequently, $M$ becomes more powerful in detecting covariate dependence of $\bmu$ or $\bgamma$.

\begin{table}[htp]
\caption{Rejection rates associated with different tests for testing $H_0^{(\bV)}$,  $H_0^{(\bmu)}$, and $H_0^{(\bgamma)}$ at nominal level 0.05}
\centering
\begin{tabular}{*{12}{c}}
\hline 
$\{n\}$ & 200 & 400 & 800 && 200 & 400 & 800 && 200 & 400 & 800 \\
\hline
\multicolumn{12}{c}{Testing $H^{(\bV)}_0: \, \bmu = \balpha_0 +\balpha_1 X, \, \bgamma = \bzero$} \\
 & \multicolumn{3}{c}{RoC} && \multicolumn{3}{c}{$M$} && \multicolumn{3}{c}{LR} \\
\cline{2-4} \cline{6-8} \cline{10-12} 
$\color{red}r$ & \multicolumn{11}{c}{$\text{DGM}^{(\bV)}_1: \, \bmu = \balpha_0^* +\balpha_1^* X, \, \bgamma = \bbeta_{0{\color{red}{r}}}^*$} \\
\cline{2-12}
0.1  & 0.075 & 0.110 & 0.160   && 0.095 & 0.105 & 0.145 && 0.070  & 0.100 & 0.220 \\
0.2  & 0.145 & 0.245 & 0.515   && 0.145 & 0.280 & 0.435 && 0.145  & 0.305 & 0.615  \\
0.4  & 0.440    & 0.870 & 0.995   && 0.360  & 0.745 & 0.935 && 0.470     & 0.900 & 0.995  \\
\hline
$\color{red}r$ & \multicolumn{11}{c}{$\text{DGM}^{(\bV)}_2: \, \bmu = \balpha_0^* +\balpha_1^* X, \, \bgamma = \bbeta_{0{\color{red}{r}}}^* + \bbeta_{1{\color{red}{r}}}^*X$} \\
\cline{2-12}
0.1  & 0.210 & 0.420 & 0.795   && 0.190 & 0.390 & 0.675 && 0.215  & 0.470 & 0.850 \\
0.2  & 0.675 & 0.990 & 1.000   && 0.525 & 0.890 & 1.000 && 0.710  & 1.000 & 1.000  \\
0.4  & 1.000 & 1.000 & 1.000   && 0.980 & 1.000  & 1.000 && 1.000  & 1.000 & 1.000  \\
\hline
\multicolumn{12}{c}{Testing $H^{(\bmu)}_0: \, \bmu = \balpha_0, \, \bgamma = \bbeta_0+\bbeta_1 X$} \\
& \multicolumn{3}{c}{RoC} && \multicolumn{3}{c}{$D$} && \multicolumn{3}{c}{$M$} \\
     \cline{2-4} \cline{6-8} \cline{10-12}
$\color{red}r$ & \multicolumn{11}{c}{$\text{DGM}^{(\bmu)}_1: \, \bmu=\balpha_0^*+\balpha_{1{\color{red}{r}}}^* X, \, \bgamma=\bzero$} \\
\cline{2-12}
0.5& 0.075 & 0.065 & 0.085 && 0.055 & 0.070 & 0.080  && 0.005 & 0.025 & 0.005    \\
1  & 0.200 & 0.265 & 0.405 && 0.195 & 0.240 & 0.420  && 0.005 & 0.025 & 0.065    \\
2  & 0.660 & 0.800 & 0.935 && 0.635 & 0.815 & 0.965  && 0.115 & 0.160 & 0.345    \\
\hline
$\color{red}r$ & \multicolumn{11}{c}{$\text{DGM}^{(\bmu)}_2:\, \bmu=\balpha_0^*+\balpha_{1{\color{red}{r}}}^* X, \, \bgamma=\bbeta_0^*$} \\ 
\cline{2-12}
0.5& 0.440 & 0.760 & 0.970 && 0.350 & 0.635 & 0.935  && 0.075 & 0.105 & 0.060     \\
1  & 0.935 & 0.995 & 1.000 && 0.930 & 0.995 & 1.000  && 0.205 & 0.160 & 0.135     \\
2  & 1.000 & 1.000 & 1.000 && 0.995 & 1.000 & 1.000  && 0.470 & 0.525 & 0.590    \\
\hline
$\color{red}r$ & \multicolumn{11}{c}{$\text{DGM}^{(\bmu)}_3:\, \bmu=\balpha_0^*+\balpha_{1{\color{red}{r}}}^* X, \, \bgamma=\bbeta_0^*+\bbeta_1^* X$} \\
\cline{2-12}
0.5& 0.850 & 0.975 & 1.000 && 0.745 & 0.935 & 1.000  && 0.140 & 0.145 & 0.260  \\
1  & 1.000 & 1.000 & 1.000 && 1.000 & 1.000 & 1.000  && 0.215 & 0.240 & 0.265  \\
2  & 1.000 & 1.000 & 1.000 && 1.000 & 1.000 & 1.000  && 0.775 & 0.880 & 0.935  \\
\hline
\multicolumn{12}{c}{Testing $H^{(\bgamma)}_0: \, \bmu = \balpha_0+\balpha_1 X, \, \bgamma = \bbeta_0$} \\
& \multicolumn{3}{c}{RoC} && \multicolumn{3}{c}{$D$} && \multicolumn{3}{c}{$M$} \\
     \cline{2-4} \cline{6-8} \cline{10-12}
$\color{red}r$ & \multicolumn{11}{c}{$\text{DGM}^{(\bgamma)}_1:\, \bmu=\balpha_0^*+\balpha_1^* X, \, \bgamma=\bbeta_{1{\color{red}{r}}}^* X$} \\
\cline{2-12}
0.5& 0.120 & 0.090 & 0.155  && 0.125 & 0.090 & 0.150   && 0.045 & 0.035 & 0.080 \\
1  & 0.135 & 0.130 & 0.185  && 0.135 & 0.125 & 0.185   && 0.030 & 0.070 & 0.065 \\
2  & 0.145 & 0.170 & 0.305  && 0.155 & 0.160 & 0.305   && 0.070 & 0.090 & 0.050   \\
\hline
$\color{red}r$ & \multicolumn{11}{c}{$\text{DGM}^{(\bgamma)}_2:\, \bmu=\balpha_0^*+\balpha_1^* X, \, \bgamma=\bbeta_0^*+\bbeta_{1{\color{red}{r}}}^* X$} \\
\cline{2-12}
0.5& 0.045 & 0.055 & 0.080  && 0.050 & 0.050 & 0.085  && 0.070 & 0.075 & 0.070  \\
1  & 0.105 & 0.095 & 0.160  && 0.105 & 0.090 & 0.160  && 0.060 & 0.055 & 0.045   \\
2  & 0.165 & 0.340 & 0.620  && 0.165 & 0.325 & 0.605  && 0.065 & 0.045 & 0.045   \\
\hline
\end{tabular}
\label{table:dependXpower}
\end{table}

\section{Prediction Regions}
\label{sec:predict}
Following the estimation of all model parameters in an ESAG regression model, one can 
predict the outcome of the directional response $\bY$. 
If all model parameters are known, similar to the prediction region for a multivariate Gaussian distribution \citep{chew1966confidence}, a sensible $100(1-a)\%$  prediction region that reflects the elliptical symmetry of ESAG$(\bmu, \bgamma)$ is an ellipsoidal ball given by  
\begin{equation}
    \text{PR}_a = \left\{ \by \in \mathbb{S}^{d-1} : \, (\by-\bmu/\|\bmu \|)^\T \bV^{-1} (\by-\bmu/\|\bmu \|) \le q_\alpha \right\}, \label{eq:PR}
\end{equation} 
where $q_a$ is chosen such that $\text{P}(\bY \in \text{PR}_a) = 1-a$. We show in Appendix A of the Supplementary Material that  $\text{PR}_\alpha$ defined in \eqref{eq:PR} has the smallest volume in a class of ellipsoidal prediction regions centering around $\bmu/\|\bmu\|$ with the nominal coverage probability of $1-a$. 

When the model parameters are unknown, we evaluate $\bmu$ and $\bV$ at their MLEs,  $\hat \bmu$ and $\hat \bV$, in (\ref{eq:PR}), and estimate $q_a$ by $\hat q_a$ that is obtained using bootstrap samples from the estimated ESAG distribution. This leads to a $100(1-a)\%$ prediction region defined as 
\begin{equation}
\widehat{\text{PR}}_a = \left\{ \by \in \mathbb{S}^{d-1} : \, (\by - \hat\bmu/\|\hat\bmu \|)^\T \hat{\bV}^{-1} (\by - \hat\bmu/\|\hat\bmu \|) \leq \hat q_a \right\}. \label{eq:PR_hat}
\end{equation}
Algorithm~\ref{alg:Prediction Region} below provides the detailed computational path leading to $\widehat{\text{PR}}_a$ when $\bX=\bx_0$. Appendix B of the Supplementary Material presents a simulation study where we follow Algorithm~\ref{alg:Prediction Region} to compute prediction regions of different nominal coverage probabilities based on samples of size $n\in \{200, 400, 800\}$. The simulation results suggest that the empirical coverage probabilities of the resultant prediction regions match closely with the nominal levels. 
\begin{algorithm}[h!]
\footnotesize
  \caption{Compute the prediction region in \eqref{eq:PR_hat}\label{alg:Prediction Region}}
  \begin{algorithmic}[1]
    \Procedure{Parametric bootstrap accounting  for variation of ESAG}{}
        \State Given the observed data $\lbrace(\bY_i,\bX_i)\rbrace_{i=1}^n$, compute the MLEs for regression coefficients, $\hat{\balpha}_0$, $\hat \bA_1$, $\hat{\bbeta}_0$, and $\hat \bB_1$, assuming an ESAG model for $\bY_i$ conditioning on $\bX_i$.
        \State Compute $\hat{\bmu}=\hat \balpha_0+
        \hat\bA_1 \bx_0$ and $\hat{\bgamma}=\hat \bbeta_0+\hat \bB_1 \bx_0$, obtain the corresponding $\hat \bV$.
        \State Set $m$ = the number of bootstrap samples. Generate a random sample, $\{\bY'_j\}_{j=1}^{m}$, from  $ \mbox{ESAG} (\hat{\bmu},\hat{\bgamma})$.
        \State Compute $q_j = (\bY'_j - \hat{\bmu}/\| \hat{\bmu}\|)^\T \hat{\bV}^{-1} (\bY'_j - \hat{\bmu}/\| \hat{\bmu} \|)$, for $j = 1,..., m$. 
    \EndProcedure
    
    \Procedure{Nonparametric bootstrap accounting for variation of MLEs}{}
        \State Set $B$ = the number of bootstrap samples.
        \For{$b$ in $1,...,B$}
        \State Generate the $b$-th bootstrap sample $\{\bY_{i}^{(b)}, \bX_{i}^{(b)}\}_{i=1}^n$ via sampling with replacement from the raw data.
        \State Repeat Steps 2--5 using data $\lbrace(\bY_i^{(b)}, \bX_i^{(b)})\rbrace_{i=1}^n$. Denote the bootstrap version of $q_j$ as $q^{(b)}_j$. 
      \EndFor 
      \State Viewing $\{q_j, \, q^{(1)}_j, \ldots , q^{(B)}_j\}_{j=1}^{m}$ as a sample of size $m\times(B+1)$, find the $(1-a)$-quantile of this sample. Denote this sample quantile as $\hat q_a$.
      \State  Output a $100(1-a)\%$ prediction region when $\bX=\bx_0$ given by $ \lbrace \by \in \mathbb{S}^{d-1} : (\by - \hat{\bmu}/\|\hat{\bmu} \|)^\T \hat{\bV}^{-1} (\by - \hat{\bmu}/\|\hat{\bmu} \|) \leq \hat q_a  \rbrace$ .   
      \EndProcedure
  \end{algorithmic}
\end{algorithm}

\section{Real-life data applications}
\label{sec:realdata}
We now put into action the regression analysis toolkit on data examples from two real-life applications.  

\subsection{Hydrochemical data}
We analyzed in a recent work \citep{yu2022elliptically} the relative abundance of two major ions, $\mbox{K}^+$ and $\mbox{Na}^+$, and two minor ions, $\mbox{Ca}^{2+}$ and $\mbox{Mg}^{2+}$, in water samples collected from two sets of locations between the summer of 1997 and the spring of 1999: 67 samples from tributaries of Anoia and 43 samples from tributaries of the lower Llobregat course in Spain \citep{otero2005relative}.  The complete data are available in the \texttt{R} package, \texttt{compositions} \citep{van2008compositions}. The relative abundance of $(\mbox{K}^+, \mbox{ Na}^+, \mbox{ Ca}^{2+},  \mbox{ Mg}^{2+})$ is an example of compositional data in a 4-dimensional simplex, $\mathbb{C}^{4-1}=\{\by^* \in \mathbb{R}^4: \, \bone_4^\top \by^*=1 \text{ and } \be_j^\top \by^*\ge 0, \text{ for $j=1, \ldots, 4$}\}$, where $\bone_4$ is the $4\times 1$ vector of ones, and $\be_j$ is the unit vector with the $j$-th entry being 1. We transformed the compositional data by taking the square-root of $\by^*\in \mathbb{C}^{4-1}$ element-wise to directional data in $\mathbb{S}^{4-1}$. Previous analyses of the directional data from each set of locations suggested an adequate fit of an intercept-only ESAG model, but a poor fit for the combined data of size $n=110$ from two sets of locations. 

These earlier findings motivate a location-dependent ESAG model for all data from these locations, where we incorporate a covariate $X$ indicating locations, with $X = 0$ corresponding to tributaries of Anoia (At), and $X = 1$ representing tributaries of lower Llobregat course (LLt). Fitting the regression model, $Y_i|X_i\sim \text{ESAG}(\bmu_i=\balpha_0+\balpha_1 X_i, \, \bgamma_i=\bbeta_0+\bbeta_1 X_i)$, to the data, we arrive at the following estimates for the ESAG model parameters, 
$$
\hat\bmu_i = \begin{bmatrix} 1.99 \\ 5.74 \\ 7.95 \\ 4.59  \end{bmatrix} + \begin{bmatrix} 1.28 \\ 2.83 \\ 1.06 \\ 1.20  \end{bmatrix} X_i,
\hspace{1cm}
\hat \bgamma_i = \begin{bmatrix} -0.67 \\ 0.15 \\ -0.82 \\ 6.12 \\ 0.64  \end{bmatrix} + \begin{bmatrix} 2.43 \\ -0.22 \\ 10.17 \\ -20.19 \\ 0.47  \end{bmatrix} X_i.
$$
Hence, for the directional response associated with tributaries of Anoia, the mean direction is estimated to be $\hat \bmu_{\text{At}} = (1.99, 5.74, 7.95, 4.59)^\top$, and, for the directional response coming from tributaries of lower Llobregat course,  the estimated mean direction is $\hat \bmu_{\text{LLt}} = (3.27, 8.57, 9.01, 5.79)^\top$. These estimates lead to the estimated concentration at each set of locations, which suggests that the latter set of locations exhibits a higher concentration than the former. These are coherent with results in our previous analysis when we analyzed data from one set of locations at a time. Estimates for $\bgamma_i$ when $X_i=0$ and 1 are also aligned with our earlier analyses (and are omitted here), based on which estimates of $\bV$ for two sets of locations, $\hat \bV_{\text{At}}$ and $\hat \bV_{\text{LLt}}$, can be obtained.

For model diagnosis, we carry out tests for isotropy and covariate dependence of $\bmu$ and $\bgamma$ based on the three proposed test statistics. All tests suggest statistically significant evidence of location-dependent model parameters in the ESAG distribution that is anisotropic for the (transformed) compositions of ($\text{K}^+, \, \text{Na}^+, \, \text{Ca}^{2+}, \, \text{Mg}^{2+}$), with all estimated $p$-values less than $10^{-3}$ except for that associated with $M$ when testing covariate dependence of $\bgamma$, which returns an estimated $p$-value less than 0.01 (although larger than $10^{-3}$). This is consistent with findings in existing literature reporting that the hydrochemical profile of  Anoia and that of the Llobergat lower course are substantially different because the two sets of tributaries pass through zones that are differently populated with vastly different distributions of agricultural and industrial areas \citep{gonzalez2012presence}. Looking more closely at the test statistics when testing covariate dependence of $\bmu$, i.e., testing the null $H_0^{(\bmu)}$, we have $D=1.062$ that is somewhat higher than $\text{RoC}=1.059$. This can be data evidence indicating that not only the norm of the mean direction depends on $X$, that is, the concentration varies across locations, but also the orientation of the mean direction differs between locations. When testing covariate dependence of $\bgamma$, i.e., testing $H_0^{(\bgamma)}$ versus the full model, the two statistics are nearly equal (at around 1.090). This suggests that, once we acknowledge a location-dependent $\bmu$ in the null model, allowing $\bgamma$ to depend on $X$ in the alternative model mostly helps to distinguish the variability of data across different locations but it may not contribute to capturing the discrepancy in the orientation of $\bmu$ in different locations. 

Lastly, applying Algorithm~\ref{alg:Prediction Region} for $x_0=0$ and 1, we obtain the prediction regions for the two sets of locations given by 
\begin{align*}
\widehat{\text{PR}}^{(\text{At})}_a & = \lbrace \by \in \mathbb{S}^3 : \, (\by - \hat\bmu_{\text{At}}/\|\hat\bmu_{\text{At}} \|)^\T \hat{\bV}_{\text{At}}^{-1} (\by - \hat\bmu_{\text{At}}/\|\hat\bmu_{\text{At}} \|) \leq \hat q^{(\text{At})}_a \rbrace, \\
\widehat{\text{PR}}^{(\text{LLt})}_a & = \lbrace \by \in \mathbb{S}^3: \, (\by - \hat\bmu_{\text{LLt}}/\|\hat\bmu_{\text{LLt}} \|)^\T \hat{\bV}_{\text{LLt}}^{-1} (\by - \hat\bmu_{\text{LLt}}/\|\hat\bmu_{\text{LLt}} \|) \leq \hat q^{(\text{LLt})}_a \rbrace,
\end{align*}
with the estimated $(1-a)$-quantiles given by  $\hat q^{(\text{At})}_a=0.029, 0.036, 0.050$ and $\hat q^{(\text{LLt})}_a=0.018, 0.023, 0.031$, for $a=0.7, 0.8, 0.9$, respectively. At each considered nominal level, having $\hat q^{(\text{LLt})}_a<\hat q^{(\text{At})}_a$  is in line with the finding that the distribution of directional data from the lower Llobregat course exhibits a higher concentration (i.e., lower variability) than that for Anoia.

\subsection{Microbiome data}
We now turn to a dataset regarding the gut microbiota of elderly adults. Besides gut microbiome compositions of 160 elderly adults, also recorded in this data include the residence types, age, body mass index (BMI), diet, and gender. A similar dataset has been analyzed by \cite{claesson2012gut}, where the authors carried out a principal component analysis to study correlations of the relative abundance of various microorganisms in the gut.   \cite{shen2022sparse} used the Gaussian chain graph model for the data to infer the effects of one's diet and residence type on gut microbiome composition. For illustration purposes, we study the potential association between two covariates, one's age and BMI, and the directional response on $\mathbb{S}^3$ defined as the square root of the relative abundance of four genera of bacteria found in the gut: Blautia, Caloramator, Clostridium, and Faecalibacterium. 

We first fit the directional response data to the ESAG regression model, for $i=1, \ldots, 160$, 
\begin{equation} 
\label{eq:ageBMI}
\bY_i|(\text{Age}_i, \text{BMI}_i) \sim \text{ESAG}(\bmu_i = \balpha_0 + \balpha_1 \text{Age}_i + \balpha_2 \text{BMI}_i, \, \bgamma_i = \bbeta_0 + \bbeta_1 \text{Age}_i + \bbeta_2 \text{BMI}_i),
\end{equation}
where $\bY_i=(Y_{i,1}, Y_{i,2}, Y_{i,3}, Y_{i,4})^\top$, with $Y_{i,j}$ equal to the squared root of the relative abundance of Blautia, Caloramator, Clostridium, and Faecalibacterium, for $j=1, 2, 3, 4$, respectively, for subject $i$, $\text{Age}_i = (\text{subject $i$'s age}- \text{the youngest subject's age}) / (\text{age range})+1$, and $\text{BMI}_i$ is similarly computed by standardizing the BMI data. Maximum likelihood estimation yields
\begin{equation}
\label{eq:estregcoef}
\begin{aligned}
    \hat \bmu_i = &\ 
    \begin{bmatrix}
        1.76 \\
        0.62 \\
        5.27 \\
        3.23 
    \end{bmatrix}+
    \begin{bmatrix}
        -1.46 \\
        0.59 \\
        -2.70 \\
        -2.93
    \end{bmatrix}
    \text{Age}_i+
    \begin{bmatrix}
        1.11 \\ 
        -0.63 \\
        1.63 \\
        2.39
    \end{bmatrix}
    \text{BMI}_i, \\
    \hat \bgamma_i = &\ 
    \begin{bmatrix}
        -6.39 \\
        0.31 \\
        3.12 \\ 
        0.63 \\
        0.69
    \end{bmatrix}+ 
    \begin{bmatrix}
        24.08 \\
        2.54 \\
        -4.29 \\
        -3.31 \\
        0.61
    \end{bmatrix}
    \text{Age}_i+
    \begin{bmatrix}
        -21.15 \\
        -2.35 \\ 
        3.07 \\
        3.20 \\
        -0.79
    \end{bmatrix}
    \text{BMI}_i.
\end{aligned}
\end{equation}

\begin{figure}[h!]
\centering
\includegraphics[width=5.5in,height=2.5in]{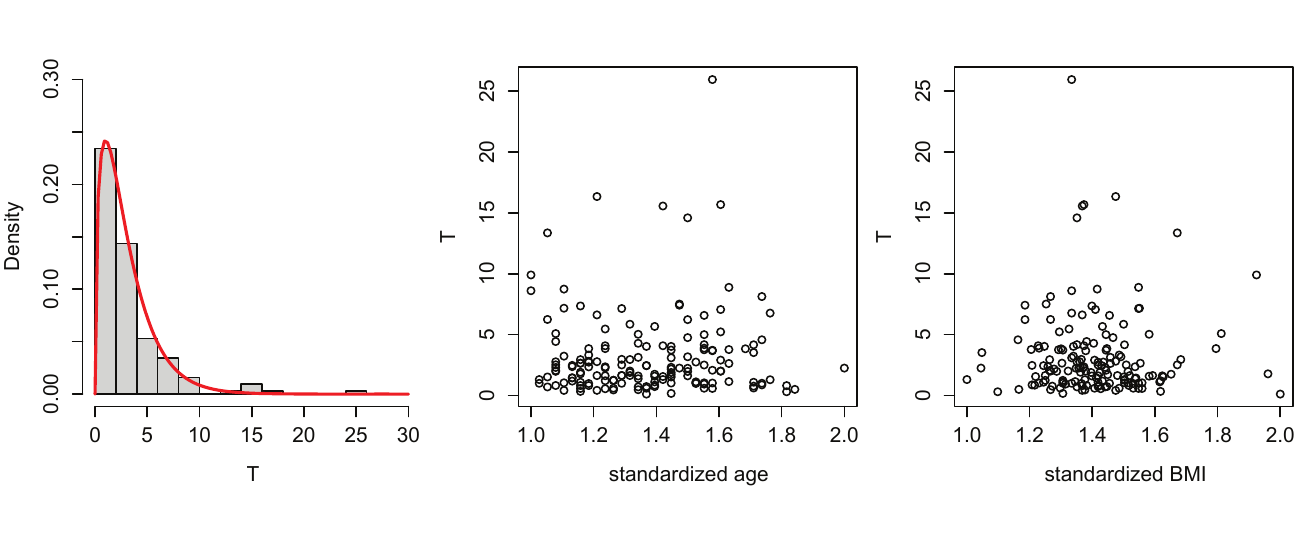}
\caption{The histogram of $\{\hat T_i\}_{i=1}^{160}$ with the density of  $\chi^2_3$ superimposed (in the left panel), the scatter plot of $\{(\text{Age}_i, \, \hat T_i)\}_{i=1}^{160}$ (in the middle panel), and the scatter plot of $\{(\text{BMI}_i, \, \hat T_i)\}_{i=1}^{160}$ (in the right panel) based on microbiome data modeled by \eqref{eq:ageBMI}.
}\label{fig:gof_BMI}
\end{figure}

We first carry out the residual-based goodness-of-fit test proposed in an earlier work \citep{yu2022elliptically}. We showed there that, if $\bY\sim \text{ESAG}(\bmu, \bgamma)$, then $T=(\|\bmu\|^2+\sum_{j=1}^d \lambda_j)\br \bV^{-1}\br$ follows $\chi^2_{d-1}$ approximately, where $\lambda_1\le \lambda_2\le \ldots \le \lambda_{d-1}$ and $\lambda_d=1$ are the eigenvalues of $\bV$, and $\br=(\bI_d-\hat \bY \hat \bY^\T)\bY$ is the directional residual \citep{jupp1988residuals} associated with the prediction $\hat \bY=\hat \bmu/\|\hat\bmu\|$. Figure~\ref{fig:gof_BMI} shows the residual-based quantities $T$ evaluated at the MLEs of unknown parameters, $\{\hat T_i\}_{i=1}^{160}$, where $\hat T_i=(\|\hat\bmu_i\|^2+\sum_{j=1}^3 \hat \lambda_{i,j}+1)\br_i \hat \bV_i^{-1}\br_i$. In particular, the empirical distribution of $T$ depicted by the histogram of $\{\hat T_i\}_{i=1}^{160}$ appears to resemble $\chi^2_3$, even though the scatter plots of $\{\hat T_i\}_{i=1}^{160}$ versus the covariates values seem to suggest several outliers in the sample. To estimate the null distribution of $T$ without approximating its distribution by $\chi^2_{d-1}$, the authors also developed a bootstrap-based test for assessing the adequacy of an ESAG model. This test applied to the current dataset yields an estimated $p$-value of 0.58, suggesting insufficient evidence for the lack of fit of the current model. In addition, the tests for isotropy based on RoC and $M$, and the tests for covariates dependence of ESAG model parameters based on RoC and $D$ all produce estimated $p$-values less than 0.01. We thus conclude significant covariates effects on the ESAG model parameters and recommend against opting for a regression model more parsimonious than \eqref{eq:ageBMI}.

\begin{figure}[h!]
    \centering
    \includegraphics[width=5.5in,height=1.7in]{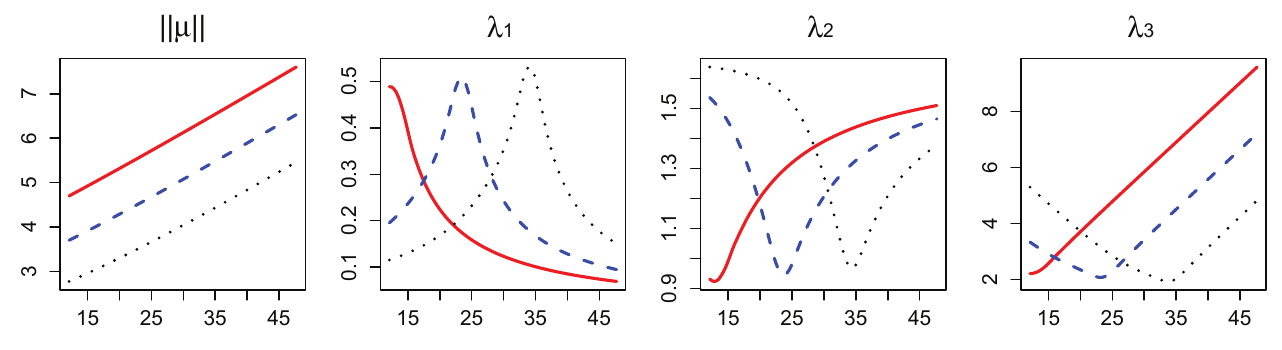}
    \caption{Estimates of $\|\bmu\|$, $\lambda_1$, $\lambda_2$, and $\lambda_3$ versus BMI when one is 70 (red solid lines), 80 (blue dashed lines), and 90 (black dotted lines) years of age.
}\label{fig:conclambda_BMI}
\end{figure}

To further elucidate the effects of age and BMI on ESAG model features, we present in Figure~\ref{fig:conclambda_BMI} estimates of the concentration and three eigenvalues of $\bV$, ($\lambda_1, \lambda_2, \lambda_3$), versus BMI when one is 70, 80, and 90 years of age. As age increases, we observe in Figure~\ref{fig:conclambda_BMI} a decrease in the estimate of $\|\bmu\|$, corresponding to an increase in the estimated overall variation of $\bY$. The finding of highly variable directional distribution can imply highly variable in the composition of the gut microbiota among the elderly, which is a finding reported in existing literature but has been mostly stated in comparison with younger (than 65) healthy adults that are found to have a more stable composition of intestinal microorganisms \citep{claesson2012gut}. Our results here can be evidence for that, even among the elderly, the trend of higher variability in microbiome composition as one ages persists. In addition, a higher BMI also leads to a more variable distribution. Examining the estimated eigenvalues of $\bV$, one can see two change points in BMI: one at BMI of nearly 25 for an 80-year-old and the other at BMI of around 35 for a 90-year-old. The first change point separates healthy weight ($\text{BMI}\in (18.5, 24.9)$) and overweight ($\text{BMI}\in (25.0, 29.9)$); the second change point belongs to the obese range (\url{https://www.cdc.gov/healthyweight/assessing}). Because $\lambda_1=\lambda_2=\lambda_3=1(=\lambda_4)$ implies $\bV=\bI_4$, the proximity of the three considered eigenvalues to 1 implies isotropy of the directional distribution and also relates to the correlations between the four genera of bacteria. The aforementioned change points are where the estimates for these eigenvalues are closest to 1, and thus the distribution of $\bY$ tends to be more isotropic when $\bX=(\text{Age}=80, \text{BMI}\approx 25)$ and $(\text{Age}=90, \text{BMI}\approx 35)$. This can also imply a reduction in the correlation between the relative abundance of the four considered genera of bacteria at these change points. 

\section{Discussion}
\label{sec:discuss}
We develop in this study a complete package of regression analysis for directional response built upon the ESAG distribution family indexed by constraint-free parameters. We consider a full range of statistical inference problems, including parameter estimation, testing hypotheses on model features, and prediction. The uncertainty of parameter estimation can be assessed via bootstrap. Parametric bootstrap is also heavily involved in all proposed inference procedures, which is straightforward to implement owing to the formulation and parametrization of ESAG that allow for easy data generation from an ESAG distribution. Computer programs for implementing all proposed methods are available at \url{https://github.com/Zehaoyu217/ESAG/blob/main/ESAG_Project2.R}. We also demonstrate the use of this package for analyzing two datasets from different fields of applications. 

The number of parameters in an ESAG regression model can be large in an application since the dimensional of the parameter space grows quadratically in the dimension of $\bY$, $d$,  and linearly in the number of covariates, $q$. For example, in microbiome analysis, $d$ is the dimension of the compositional response, which  typically is much larger than four, and one may wish to consider many covariates relating to 
 the host's physiological characteristics. We have started developing penalized likelihood-based methods to deal with high-dimensional directional data. Besides this ongoing follow-up research, another interesting topic is compositional data analysis that the two case studies in Section~\ref{sec:realdata} relate to. The idea of relating compositional data on a simplex to directional data on a hypersphere has been explored \citep{scealy2011regression, scealy2017directional, li2023reproducing} but with many open questions yet to be addressed. In this particular context, more components may have zero or nearly zero relative abundance as $d$ increases, which is a data pattern ESAG and most existing named directional distributions tend to fit poorly. Interpretations and implications of model parameters of a directional distribution that are practically meaningful for the corresponding compositional data also demand further systematic investigation.



\clearpage
\bibliographystyle{apalike}
\bibliography{ref.bib}

\appendix

\renewcommand*{\theequation}{A.\arabic{equation}}
\setcounter{equation}{0}
\setcounter{lemma}{1}
\section*{Web Appendix A: Prediction regions of the smallest volume}
Suppose $\bY\sim \text{ESAG}(\bmu_0, \bgamma_0)$ resulting from normalizing $\bW\sim \mathcal{N}_d(\bmu_0, \bV_0)$. Define a class of $100(1-a)\%$ prediction regions $\{\text{PR}_a(\bV):\, \bV \text{ is a $d\times d$ positive definite matrix satisfying } \bV\bmu_0 = \bmu_0 \text{ and } \text{det}(\bV) =1\}$, where 
\begin{equation}
    \text{PR}_a (\bV)= \left\{\by \in \mathbb{S}^{d-1} : \, (\by-\bmu_0/\|\bmu_0 \|)^\top \bV^{-1} (\by-\bmu_0/\|\bmu_0 \|) \le q_a \right\}, 
\end{equation} 
with $q_\alpha$ chosen such that $P(\bY \in \text{PR}_a(\bV)) = 1-a$. In what follows, we show that $\text{PR}_a (\bV_0)$ has the smallest volume in this class.

Because 
\begin{align*}
    & (\bY-\bmu_0/\|\bmu_0\|)^\top \bV^{-1} (\bY-\bmu_0/\|\bmu_0 \|) \\
    = & \bY^\top\bV^{-1}\bY - 2\frac{\bmu_0^\top}{\|\bmu_0\|}\bV^{-1}\bY + \frac{\bmu_0^\top}
    {\|\bmu_0\|}\bV^{-1} \frac{\bmu_0}{\|\bmu_0\|} \\
    = & \bY^\top\bV^{-1}\bY - 2\frac{\bmu_0^\top}{\|\bmu_0\|}\bY + 1, \text{ since $\bmu_0^\top \bV^{-1}=\bmu_0^\top$,}
\end{align*}
which depends on $\bV$ only via $\bY^\T\bV^{-1}\bY$. We further elaborate on this term next. With $\bY=\bW/\|\bW\|$, we re-express $\bY$ as $(\bV_0^{1/2}\bZ+\bmu_0)/\|\bW\|$, where $\bZ \sim \mathcal{N}_d(\bzero, \bI_d)$. It follows that 
\begin{align*}
    & \bY^\top\bV^{-1}\bY \\
    = &\ \frac{\bZ^\top\bV_0^{1/2}\bV^{-1}\bV_0^{1/2}\bZ + 2\bmu_0^\top \bV^{-1}\bV_0^{1/2}\bZ+\bmu_0^\top\bV^{-1}\bmu_0}{\|\bW\|^2} \\
    = &\ \frac{\bZ^\top\bV_0^{1/2}\bV^{-1}\bV_0^{1/2}\bZ + 2\bmu_0^\top \bV_0^{1/2}\bZ+\|\bmu_0\|^2}{\|\bW\|^2},
\end{align*}
which depends on $\bV$ only via $\bZ^\T\bV_0^{1/2}\bV^{-1}\bV_0^{1/2}\bZ=\bZ^\top \bG \bZ$, where $\bG = \bV_0^{1/2}\bV^{-1}\bV_0^{1/2}$. Because $\det(\bV_0)=\det(\bV)=1$, $\bG$ is a symmetric positive definite matrix with a determinant equal to one.   

Note that $q_a$ is affected by $\bV$ via the variation of $\bZ^\top\bG\bZ$. To show that $\text{PR}_a(\bV_0)$ has the smallest volume in the class of prediction regions defined above, it suffices to show that $\text{Var}(\bZ^\top\bZ) \leq \text{Var}(\bZ^\top\bG\bZ)$. Because $\bZ^\top \bZ\sim \chi^2_d$, $\text{Var}(\bZ^\top\bZ) = 2d$. Using the eigendecomposition of $\bG$ given by $\bP\bD \bP^\top$, we have 
\begin{align*}
    \text{Var}(\bZ^\top\bG\bZ) & = \text{Var}(\bZ^\top\bP\bD\bP^\top\bZ) \\
    & = \text{Var}(\bU^\top\bD\bU), \ \ \ \mbox{where}\ \bU =(U_1, \ldots, U_d)^\top= \bP\bZ \sim \mathcal{N}_d(\bzero,\bI_d), \\ 
    & = \sum_{j=1}^d D_j^2\text{Var}(U_j^2), \text{ where $D_j$ is the $j$-th diagonal entry of $\bD$}, \\ 
    & = 2\sum_{j=1}^d D_j^2, \text{ since $U_j^2\sim \chi_1^2$, for $j=1, \ldots, d,$}
\end{align*}
where $D_j>0$, for $j = 1,...,d$, and $\prod_{j=1}^d D_j = \det (\bG)=1$. By the arithmetic–geometric mean inequality, 
$$ \frac{1}{d}\sum_{j=1}^d D_j^2 \ge \left( \prod_{j=1}^d D_j^2\right)^{1/d},$$
with the right-hand side equal to one since $\prod_{j=1}^d D_j^2=\prod_{j=1}^d D_j=1$. Hence, $\sum_{j=1}^d D_j^2 \ge d$, and thus $\text{Var}(\bZ^\top\bG\bZ)\ge \text{Var}(\bZ^\top\bZ)$.

In conclusion, setting $\bV = \bV_0$ so that $\bG=\bI_d$ results in the lowest variation in $(\bY-\bmu_0/\|\bmu_0\|)^\top \bV^{-1} (\bY-\bmu_0/\|\bmu_0\|)$, which leads to the smallest value of $q_a$, and further leads to the smallest volume for the prediction region. 

\section*{Web Appendix B: Simulation study on prediction regions}
For illustration purposes, we carry out a simulation study where we implement Algorithm~4 to construct prediction regions based on random samples of size $n\in \{200, 400, 800\}$ from an ESAG model with intercept-only models for both $\bmu$ and $\bgamma$, with $\bmu = (2,-5,3,5)^\top$ and $\bgamma = (3,5,-3,4,2)^\top$. Following constructing a prediction region based on a random sample, we compute the proportion of the sample falling in this region, viewed as an empirical coverage probability of the prediction region. 

Table~\ref{table:coverPR} provides Monte Carlo averages of the empirical coverage probabilities across 2000 Monte Carlo replicates at each simulation setting specified by the level of $n$. These summary statistics suggest satisfactory performance of the proposed prediction region in that they achieve the desired coverage probabilities even when the sample size is moderate.
\begin{table}[h]
\begin{center}
\caption{\label{table:coverPR}Monte Carlo averages of empirical coverage probabilities associated with prediction regions at three nominal levels $1-a$. Numbers in parentheses are Monte Carlo standard errors of corresponding averages.}
\begin{tabular}{cccc}
\hline
 $1-a$ & $n = 200$         & $n = 400$        & $n = 800$        \\ \hline
0.90 & 0.895 (0.0174) & 0.897 (0.0121) & 0.898 (0.0089) \\
0.95 & 0.947 (0.0111)  & 0.948 (0.0080)  & 0.949 (0.0058) \\
0.99 & 0.989 (0.0350)   & 0.989 (0.0026) & 0.990 (0.0019) \\
\hline
\end{tabular}
\end{center}
\end{table}

\end{document}